# A low-cost indoor and outdoor terrestrial autonomous navigation model

Gianluca Susi, *Member, IEEE,* Alessandro Cristini, Mario Salerno, and Emiliano Daddario

*Abstract* — **In this paper, a method for low-cost system design oriented to indoor and outdoor autonomous navigation is illustrated. In order to provide a motivation for the solution here presented, a brief discussion of the typical drawbacks of state-of-the-art technologies is reported. Finally, an application of such a method for the design of a navigation system for blindfolded people is shown.**

## I. Introduction

THERE are a lot of systems proposed in the literature that face the issue of autonomous navigation, applied for several purposes, such as guidance of land vehicles [1]–[6], UAVs (Unmanned Aerial Vehicles) [7], [8], robotic applications [9]–[12], and also in order to improve the quality of life of visually impaired people [13]–[16]. The methods proposed are based on different technologies, such as GPS (Global Positioning System) [7], IR (Infrared) [9], [13], RFID (Radio Frequency Identification) [5], [10], [16], Ultrasonic [12], [16], and combinations of various technologies, including GPS-INS (Inertial Navigation System) [1]–[4], IR-Ultrasonic [14], RFID-Ultrasonic [6], [11], Ultrasonic-INS [8], etc.

However, these technologies present some drawbacks. For example, the accuracy of GPS is not reliable when operating indoors due to limited satellite reception [10], moreover, today's GPS based navigation systems often do not function well at cities with tall buildings [5]; Ultrasonic cannot provide high angular resolution due to the wide beam angle (which can be about 30 degrees or even wider) [8]; RFID systems are generally expensive and attempting to read several tags at a time may result in signal collision and ultimately to data loss [17]; furthermore, RFID readers cannot be installed at every location [18]. Some limitations can be overcome using systems based on sensor fusion (e.g., IR-Ultrasonic, RFID-Ultrasonic, etc.), but that involves complex and often expensive systems.

Corresponding author: Gianluca Susi is with the Department of Electronic Engineering, University of Rome "Tor Vergata", Via del Politecnico 1, 00133 Rome, Italy (e-mail: gianluca.susi@uniroma2.it).

Alessandro Cristini is with the Department of Electronic Engineering, University of Rome "Tor Vergata", Via del Politecnico 1, 00133 Rome, Italy (e-mail: alessandro.cristini@students.uniroma2.eu).

Mario Salerno is with the Department of Electronic Engineering, University of Rome "Tor Vergata", Via del Politecnico 1, 00133, Rome, Italy (salerno@uniroma2.it).

Emiliano Daddario is with the Department of Electronic Engineering, University of Rome "Tor Vergata", Via del Politecnico 1, 00133, Rome, Italy (emiliano182@yahoo.it).

A typical autonomous navigation system could be outlined as follows:
1. A sensing unit (SU, e.g., depth camera, ultrasonic, etc.);
2. A processing unit (PU);
3. A feedback/control unit (FCU, e.g., actuators, displays, headphones, etc.).

In addition, in order to realize a portable system, usable even in outdoor scenarios, an on-board power supply should be used.

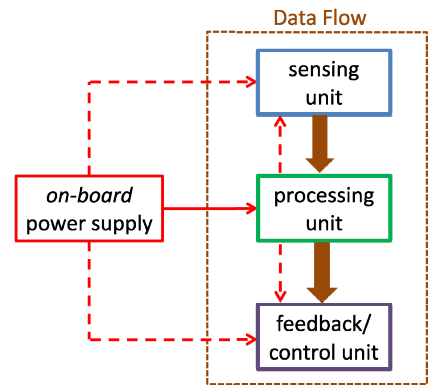

Fig. 1. Scheme of a typical autonomous navigation portable system.

In the last years, many works have been proposed in the field of autonomous navigation, in which *depth cameras* are used as *sensing units* [19]–[21].

In this work, a low-cost model for indoor and outdoor terrestrial navigation, based on depth data, will be presented. We will introduce today's available low-cost technologies, and then we will present a method to be used in the design phase, making possible to avoid typical problems arising from the technologies taken into account.

Finally, a prototype aimed at human blind autonomous navigation will be presented, and the results will be shown in terms of performance and cost.

## II. Structured Light and Time-of-Flight consumer depth cameras

As a newly developing category of distance measuring hardware, the depth camera technologies opened a new epoch for 3D geometric information acquisition. A plenty of applications like geometry reconstruction, mixed reality and human motion tracking, can be obtained thanks to the fast analysis of 3D scenes. Until a few years ago, the cost of a depth camera was approximately 10 k€; from 2010 the SL (Structured Light) depth sensing technology is available on the large-scale market, and then many products based on this technology have been released

(e.g., Microsoft Kinect v.1, Asus Xtion, etc.). The subsequent cost reduction (approximately 0.1 k$) together with their sensing range and compatibility, have allowed the use of RGB-D (RGB-Depth) input devices in many everyday applications, such as gaming, biomedical field, art and many others. Also, there has been a remarkable widespread use of low-cost sensing technologies on interesting navigation tasks (e.g., indoor blind navigation, underground exploration, autonomous mapping of buildings, night vision, or remote exploration of dangerous areas [22] ).

Limitations of these kinds of approach have been stressed in the literature [23], and some of these can lead to specific problems (e.g. *disparity holes* for transparent, shiny or matte and absorbing objects, sunlight interference) [24]–[25]. These drawbacks can be reduced by proper techniques and algorithms, which usually exploit the large amount of information that it is possible to obtain through the devices by data fusion (e.g., multiple IR camera configurations and RGB-Depth data fusion) [26], [27], [28].

In these years, consumer depth-cameras based on other technologies have appeared on the consumer market. In particular, ToF (Time-of-Flight) technology (e.g., Senz3d, Kinect v2.0) is becoming affordable (less than 0.2 k$). Furthermore they are able to show better performances in some scenarios: higher SNR and resolution, higher sensing range, no depth shadow due to the single viewpoint. In recent models, a better response to varying lighting conditions is provided thanks to the integration of ambient light rejection filters. ToF cameras are the new attractive candidates to be used in low-cost depth sensing based systems.

III. THE MODEL

In this section a low-cost model is introduced; despite the presence of the mentioned drawbacks, the aim of this work will be to preserve the reliability for navigation tasks. For this purpose, different design choices will be identified, in regard to the specific application.

*A. Hardware*

The proposed model is shown in Fig. 2.
With the aim of schematizing the information/power flow, the system will be described as a particularized version of the general one as follows. It has been depicted as a set of seven *blocks* properly connected: *sensing b.* (block 1), *processing b.* (block 2), *media encoding b.* (block 3), *actuator driver b.* (block 4), *media output b.* (block 5), *actuator b.* (block 6), *on-board power supply b.* (block 7).

Block 1 is the root of the information flow; it consists of a SL or ToF RGB-D input device, allowing depth mapping in absence of light. For outdoor navigation, because of the sunlight interference, a new-generation consumer ToF camera is suggested (as the Kinect v.2 that implements an integrated ambient light rejection filter). It is directly connected to block 2. Because of the high data throughput generated by this CV (Computer Vision)-based system, a wired connector is needed between them. One of the most common connector types is USB 2.0, or 3.0 (e.g., Kinect v.2 for Windows).

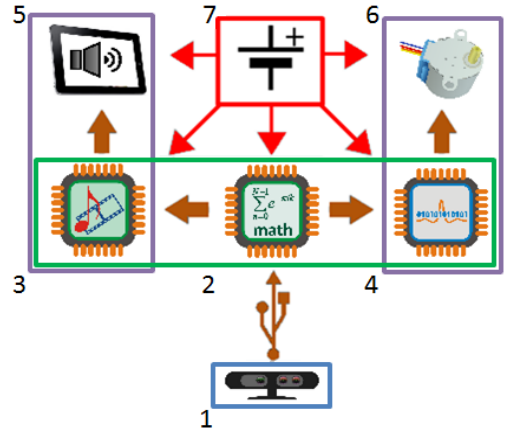

Fig. 2. Hardware scheme of the navigation aid model; colors are related to that in the previous figure.

Block 2 should consist of a small embedded Linux computer, for high performance computing and increased portability. This block numerically processes the input stream through the *processing chain* (explained in the next subsection), in order to compute the current internal status of the system, and then it controls block 3 and block 4 accordingly. These can either be featured inside the same Linux computer, or consist of external boards (e.g. block 3 can be a programmable audio DSP). Block 3 is responsible for converting the received system status into a meaningful sound/video output for the user (e.g., DAC, hardware acceleration). For maximum comfort, Bluetooth devices can be used as the media output (i.e., block 5). Block 4 is responsible for driving block 6. During the prototype phase, a GPIO (General Purpose Input/Output) interface between them is often used. Block 4 can be a different controller from block 2, for example a fast prototyping board (e.g., Arduino) connected to it via wired serial interface or wireless communication interface (Wi-Fi for easy setup, Bluetooth Low Energy for extended battery life). However, some popular embedded Linux computers include GPIO interfaces for physical computing (e.g., BeagleBone Black), then including block 2 and block 4. Block 6 is responsible of providing tactile and/or proprioceptive feedback to the end user, also acting as a haptic display used to convey visual information. In Fig. 2, the latter is represented by a stepper motor with a threaded shaft used as a linear actuator, but other solutions can be taken into account. In non human-oriented systems the *FCU* can consist of a control block connected to the subsequently device.

Block 7 guarantees portability. Of course, some blocks (such as wireless headphones for block 5) may use their own battery, especially for wireless operation; other blocks (e.g., block 1) can be powered from block 2 via USB (e.g., Xtion).

*B. Software*

The choice of a consumer RGB-D input device as sensing unit implies a lot of advantages: above all, the chance to realize a low cost implementation. However, this kind of devices presents peculiarities that can lead to specific problems when applied in navigation tasks. In order to overcome these typical problems, a *processing chain* (implemented by *block 2*) is provided.

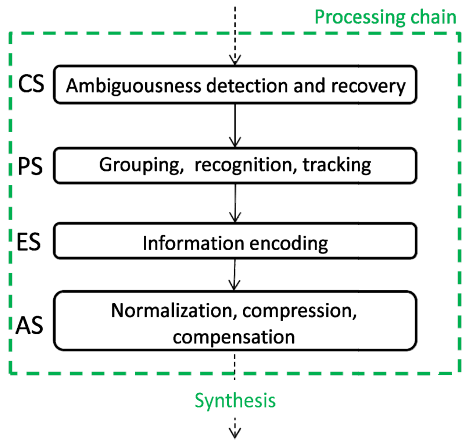

Fig. 3. The *processing chain* schematizes the steps by which the processing block elaborates the data in the autonomous navigation model proposed.

The processing chain can be divided in four steps, as follows.

The CS (Correction Step) is an elaboration block that implements algorithms aimed at recovering missing depth data, as in the case of reflective, transparent or matte surfaces, depth discontinuity, or sunlight interference. Note that, this block is crucial for the correct detection and avoidance of obstacles on a route. Many approaches can be used for limiting this problem. For example, to recover *disparity holes*, a *joint-bilateral filter* can be iteratively applied to depth pixels, through the consideration of RGB data, depth information and a temporal consistency map, as shown in [29]; to recover sunlight interference in ToF cameras the method proposed in [30], based on RGB data and depth information, can be applied. Other methods using both RGB and Depth data are reported in reference [31]; note that, the need of RGB data makes these approaches not compatible with dark condition scenarios. This limitation could be overcome by combining structured light with IR stereo techniques, as illustrated in [26]. The latter solution is very effective, but it requires a couple of devices, and then an appropriate synchronization process to avoid interference. The device overlapping is also used with the aim of extending the coverage, improving the resolution and overcoming occlusions. Another method used in order to avoid the interference among cameras consists of giving a small amount of motion to the sensors [27], [28]. With the purpose of making an appropriate choice among the solutions above mentioned, considerations about computing capabilities and, on the other hand, complexity and latencies introduced by the different approaches have to be taken into account.

The PS (Processing Step) operates the grouping of reference areas from the depth map flow. To extract information from a set of pixels many methods can be used, in relation to the specific application: a simple evaluation of the average or the maximum on sets of pixels can be made for a low level feedback information encoding; also, tracking/recognition algorithms can be introduced at this step. A review of such methods is listed in [31].

The ES (Encoding Step) converts the information into a code, that can be used for audio and/or tactile (e.g., blind people), or video (e.g., navigation in dark conditions) feedback. A proper choice of the kind of feedback to be used is necessary. For example, in some human-oriented applications, the audio feedback could result in masking phenomena on natural information coming from the environment. For instance, in order to avoid these phenomena, for the case of sightless navigation applications a tactile feedback is preferred.

The AS (Adapter Step) provides the representation of the information to the user senses. For example, it can realize the normalization/compression on the excursion of linear actuators, regulation of the amplitudes related to different audio-coded items or the compensation of human adaptation phenomena (audio/tactile memory effect).

IV. IMPLEMENTATION AND RESULTS

With the aim of evaluating the effectiveness of the model, a low-cost prototype system aimed at human blind autonomous navigation has been realized using the design principles illustrated above, and tested on a specific task.

The prototype is composed as follows: Kinect v.1 (*block 1*); Raspberry Pi (*block 2* and *block 3*); headphones (*block 5*). On the basis of the acquired depth data, the system generates proper acoustic feedbacks, synthesized by means of both an accurate analysis and the interpretation of the depth map. DSP software based on PureData for the sound engine, SimpleOpenNI and Open Sound Control OSCp5 Libraries, has been used for the *processing chain,* as described in [32], but adding a compression curve implemented as *AS,* in order to emphasize the sound amplitudes related to near objects. The system is supplied by an on board battery (*block 7*).

We have arranged four walking paths by means of proper obstacle sets placed in different ways in a closed room, preserving the same difficulty for each path as in [32]. The system synthesizes the acoustic feedback related to the obstacle detection for a given path. Sets of trials on five blindfolded people (four trials per people) have been made, and a first set of TT (Travel Times) and NoC (Number of Collisions) have been obtained. The data obtained show satisfactory results in terms of TT, i.e., an adaptation of the individuals to the proposed method. On the other hand, NoC results almost constant.

In a second phase, we have replaced the audio feedback with the tactile one, repeating the trials, providing the information by means of a belt array composed of four actuators (*block 6*) controlled by Arduino (*block 4*). Similar performance in terms of TT have been obtained, but NoC performance have been improved. In relation to four consecutive trials performed by the individuals on the different walking paths, The performance is summarized in Table 1. The total system cost is reported in Table 2.

TABLE 1: AVERAGES OF TT AND NOC FOR THE TWO CONFIGURATIONS, A (AUDIO) AND T (TACTILE).

| Conf. | Trial n.1 | Trial n.2 | Trial n.3 | Trial n.4 |
|---|---|---|---|---|
| TT(A) | 167 s | 150 s | 65 s | 68 s |
| TT(T) | 190 s | 146 s | 101 s | 80 s |
| NoC(A) | 4.5 | 5 | 3.75 | 4.25 |
| NoC(T) | 4.75 | 3.25 | 2.75 | 2 |

TABLE 2: SYSTEM COSTS

| Device | Approx. Cost ($) |
|---|---|
| Kinect v.1 | 130 |
| Headphones | 20 |
| Arduino | 60 |
| Raspberry Pi | 75 |
| Battery | 20 |
| Linear actuators + el. components | 25 |
| TOTAL | 330 |

Future improvements will consist in making the system able to work at a higher depth range, and increasing robustness to strong-light conditions. The best solution could be data fusion obtained by new generation consumer ToF RGB-D cameras (e.g., Kinect v.2). The cost of these kind of devices remain affordable. Both the plenty of applications that can be realized and the better quality of the obtainable information, result in a very attractive approach.